\documentclass[baaa]{baaa}

\usepackage[pdftex]{hyperref}
\usepackage{subfigure}
\usepackage{natbib}
\usepackage{helvet,soul}
\usepackage[font=small]{caption}

\begin{document}


\journalvol{61A}
\journalyear{2019}
\journaleditors{R. Gamen, N. Padilla, C. Parisi, F. Iglesias \& M. Sgr\'o}


\contriblanguage{1}


\contribtype{2}

\thematicarea{0}

\title{Stirring up an embedded star cluster with a moving gas filament}
\subtitle{}


\titlerunning{Stirring a cluster with moving filament}


\author{D. Matus Carrillo\inst{1}, M. Fellhauer\inst{1}, \& A. Stutz\inst{1,2}}
\authorrunning{Matus Carrillo et al.}


\contact{dimatus@udec.cl}

\institute{
Departmento de Astronom\'{i}a, Facultad de Ciencias F\'{i}sicas y Matem\'{a}ticas, Universidad de Concepci\'{o}n, Concepci\'{o}n, Chile \and
Max-Planck-Institute for Astronomy, K\"onigstuhl 17, 69117 Heidelberg, Germany\\
}


\resumen{
    Realizamos simulaciones para probar los efectos de un filamento de gas en
    movimiento sobre un {{c\'umulo}} estelar joven (i.e. el modelo {\it slignshot}).
    {{Modelamos un {\it Orion Nebula Cluster}}} como una esfera de Plummer, y el
    Filamento con Forma de Integral como un potencial cilíndrico, donde su
    posición viene dada por una función sinusoidal.
    {{Observamos que en un filamento estático}}, un {{c\'umulo}} inicialmente esférico
    evoluciona de forma natural hacia una distribución elongada de estrellas.
    {{Para un filamento en movimiento}}, observamos distintos remanentes y los
    clasificamos en 4 categorías.
    Cúmulos ``saludables'', donde casi todas las estrellas permanecen dentro del 
    filamento y del {{c\'umulo}}; cúmulos ``destruidos'' son el caso opuesto, casi sin 
    estrellas en el filamento o cerca del centro de densidad del {{c\'umulo}}; cúmulos 
    ``ejectados'', donde una fracción mayor de estrellas permanecen unidas al
    {{c\'umulo}}, pero casi ninguna  se mantiene dentro del filamento; y cúmulos de
    {{``transición''}}, donde aproximadamente el mismo {{n\'umero}} de partículas
    es eyectado del {{c\'umulo}} y del filamento.
    Un {{c\'umulo}} con las características de {\it Orion Nebula Cluster} podría
    permanecer dentro del filamento o ser eyectado, pero no ser\'a destruido.
}

\abstract{
    We perform simulations to test the effects of a moving gas filament on a young star 
    cluster (i.e. the ``Slingshot'' Model).
    We model Orion Nebula Cluster-like clusters  as Plummer spheres and the Integral
    Shaped Filament gas as a cylindrical potential. 
    We observe that in a static filament, an initially spherical cluster evolves 
    naturally into an elongated distribution of stars.
    For sinusoidal moving filaments, we observe different remnants, and classify 
    them into 4  categories.
    "Healthy" clusters, where almost all the stars stay inside the filament and the 
    cluster; "destroyed" clusters are the opposite case, with almost no particles in the 
    filament or near the centre of density of the clusters; "ejected" clusters, where a 
    large fraction of stars are close to the centre of density of the stars , but
    almost none of them in  the filament; and "transition" clusters, where roughly
    the same number of particles is ejected from the cluster and from the filament.
    An {{Orion Nebula Cluster-like}} cluster might stay inside the filament or be ejected, 
    but it will not be  destroyed.
}


\keywords{methods: numerical, Galaxy: open clusters and associations: individual (ONC)}

\maketitle
\section{Introduction}
The Orion nebula is the nearest site of massive star formation.
One of its predominant features is the Integral Shaped Filament
\citep[ISF,][]{bally_1987}, a filament of gas in where the Orion Nebula 
Cluster \citep[ONC,][]{hillenbrand_1998} is  forming.

Observations of the protostars on the ONC show  that they are distributed
in different ways in space: protostars are located right on top of the 
ridgeline of the filament, meanwhile pre-main-sequence stars  are symmetrically
distributed around the filament \citep{stutz_2016, kainulainen_2017,stutz_2018}.
A similar behaviour is observed  for the radial velocity of the stars, where
 protostars  have radial velocities close to the velocity of the gas,
with a low velocity dispersion, and the pre-main-sequence stars have a larger velocity
dispersion of the order or larger than the velocity of the gas \citep{stutz_2016}.

\begin{figure*}
    \centering
    \includegraphics[height=8cm]{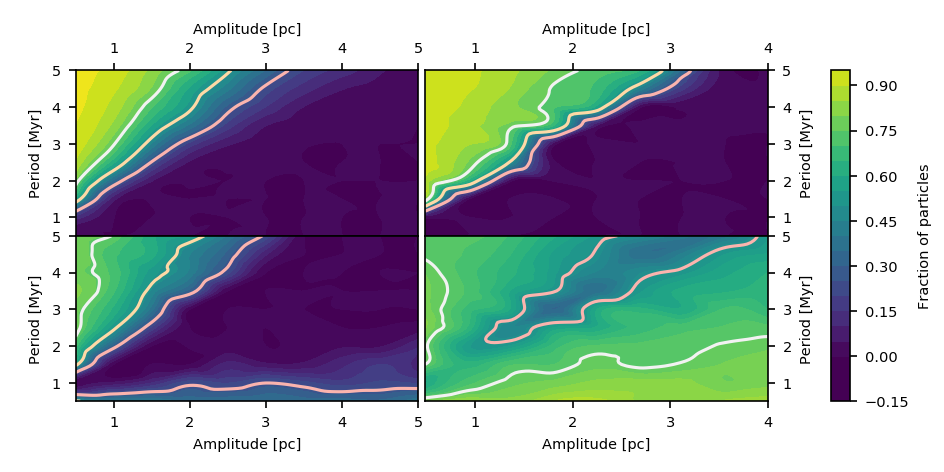}
    \caption{Fraction of particles inside the filament (top) and
    inside the cluster (bottom) after one oscillation for a cluster
    with $R_\mathrm{pl} = 0.1$~pc,  $M_\mathrm{pl} = 250$~M$_\odot$ (left) 
    and $R_\mathrm{pl} = 0.1$~pc,  $M_\mathrm{pl} = 1000$~M$_\odot$ (right).
    The contours indicate 75\%, 50\% and 25\% of the initial number of particles.
    Notice that the number of particles inside the filament drops
    drastically after it reaches $\sim50\%$ for the more massive cluster,
    but decreases smoothly for the less massive cluster.
    }
    \label{fig:fractions}
\end{figure*}

To explain these observations, \citet{stutz_2016} proposed a scenario 
where the gas of the ISF oscillates.
The protostars are formed from this gas and move with the oscillating 
filament.
Once they accrete enough mass, they are ejected from their birthplace
and start moving with the transverse velocity of the oscillation at the
moment of decoupling.
\citet{stutz_2016} named this scenario ``the Slingshot''.

To test this hypothesis, \citet{boekholt_2017} showed that an 
initially narrow distribution of stars can be dynamically heated to
produce a broad distribution with a net relative velocity  when the filament
oscillates with a certain amplitude and period. 
Moreover, Gaia data of the protostars plus the radial velocities of the gas
confirm that the gas of the ISF moves like a standing wave \citep{stutz_2018a}.

In this work we continue the exploration of the Slingshot via simulations, this
time replacing the string of stars with spherical star clusters of different masses.

\section{Method}
\subsection{The gas filament}
To study the effect of the ISF on a young star cluster, we build a cylindrical
potential as in \citet{boekholt_2017}, using the observed line density of the 
gas at the position  of the ONC \citep{stutz_2018} as a constrain to obtain the
relevant parameters of the model in the region of interest.

The ISF likely moves due to the interplay of gravitational and magnetic forces
\citep{stutz_2016,dominik_2018}, but for this work we assume that the gas  potential is
moving along the x-axis like a harmonic oscillator:
\begin{equation}
    x(t) = A\sin\left(\frac{2\pi}{P}t\right)
\end{equation}
where $A$ is the amplitude of the oscillation and $P$ is the period.

\subsection{The star cluster}
We use a cluster of equal mass point particles to represent the ONC.
These particles are distributed following a Plummer profile \citep{plummer_1911}
and located inside the cylindrical potential.

The presence of an external potential means that there is an extra component to
the net force that the particles feel, but due to the symmetry of the filament,
that extra acceleration can be non-zero only in the \mbox{x-y} plane.
A standard Plummer sphere will, then, collapse due to the extra mass of the 
filament along those axes.
We avoid that initial contraction by increasing the \mbox{x-y} velocity of the
particles so that the cylindrical Lagrangian radii of the cluster inside the 
filament remains roughly constant.

\subsection{The code}
The system of particles is evolved using the \texttt{ph4} N-body code
\citep{mcmillan_2012}. 
To account for the gas filament, we implement the background potential using
\texttt{amuse} \citep{pelupessy_2013}.
We couple this potential with the  N-body solver using the 
\texttt{bridge} \citep{fujii_2007} method, in which the velocities of 
the stars are periodically updated using the acceleration due to the filament.

\section{Results}
{
Due to the cylindrical symmetry of the potential used to represent the filament,
the embedded clusters tend to elongate along the z axis.
Particles moving throughout the length of the gas filament will do so following
corkscrew orbits around the center of the potential.
}

We determine the state of the cluster by counting the stars inside the filament,
measured as the number of stars that never move beyond $5\times R_\mathrm{pl}$ 
from the centre of the filament.
Inside the cluste we define the evolution in a  similar fashion using the centre
of density of the particle system instead.

If the filament moves slowly, a large fraction of particles stays inside the
filament and the cluster.
As the amplitude of the oscillation increases,
or the period decreases, a larger fraction of stars are ejected from the system,
up to the moment where no stars are inside the filament (Fig.
\ref{fig:fractions}, top left) or close to the centre of density of the stars 
(Fig. \ref{fig:fractions}, bottom left).
If the filament moves too quickly, it will not spend enough time inside the
cluster to perturb the particles and a large fraction of stars (but not all
of them, so we can still identify  an overdensity of particles)  will be 
ejected from the cluster.

A more massive cluster will have a similar behaviour.
However, being more tightly bound, it will not be destroyed 
(Fig. \ref{fig:fractions}, bottom right). 
The fraction of particles in the  filament will have a sharper transition from 
the high values characteristic for  the slow filaments to the empty filament 
state (Fig. \ref{fig:fractions}, top right).

\begin{figure}
    \centering
    \includegraphics[width=8cm]{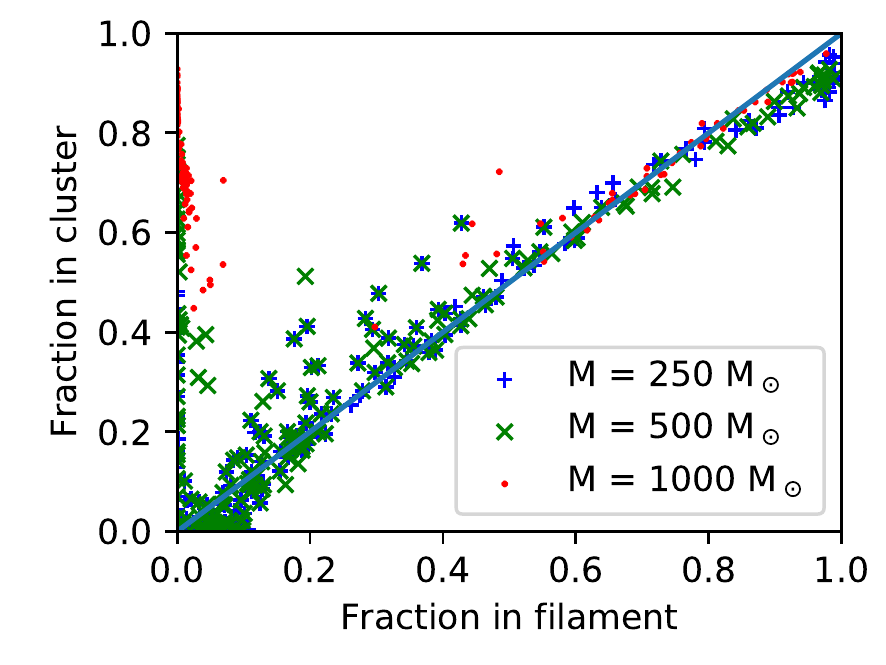}
    \caption{Fractions of particles inside cluster and filament for models with
    $R=0.1$ pc after one full oscillation.   
    The blue line is a reference for a 1:1 ratio.
    }
    \label{fig:scatter}
\end{figure}
Fig. \ref{fig:scatter} shows the fraction of stars inside the filament,
 versus the fraction of stars inside the cluster,
for the clusters with $R_\mathrm{pl} = 0.1 $~pc.
 This shows us a sequence of remnants after one full oscillation of the filament
that can be catalogued into 4 groups:
\begin{itemize}
    \item Slow filaments (i.e. large period and/or small amplitudes) will keep a large number of particles inside the cluster and the filament, and in this case, it behaves like a cluster inside a static filament. Some stars are ejected from these "healthy" clusters, since the filament is still injecting energy into the cluster. 
    \item Some combinations of period and amplitude will produce remnants that are completely destroyed, where only a small ($<$ 20 \%) fraction of stars will stay inside the filament and move with it. 
    \item A high fraction of stars in the cluster, but almost none in the filament is observed when the cluster is ejected from the filament. There is negligible mass loss when the cluster is outside the filament
    \item The last kind of objects are those where the same number of particles are ejected from the filament and from the cluster. Most of the mass loss happens between the beginning of the simulation and by the time the filament reaches its maximum amplitude. After that, there are small losses each time the filament reaches the maximum distance from the origin.
\end{itemize}

Not all the models will produce remnants in every category.
As can be seen in Fig. \ref{fig:scatter},  none of the clusters 
with  $M_\mathrm{pl} = 1000 $~M$_\odot$ are destroyed. 
Moreover there is a gap where none of the cluster remnants end with $<40\%$
of the particles inside the filament, with the exception of the ejected clusters.
This means that the cluster can stay inside the filament or it can ejected
after losing less than half of its mass, but it is not destroyed by the filament.

\section{Conclusions}
We present results of the first simulations of the effects of an oscillating
cylindrical  potential in the early evolution of a young star cluster.
We explore the space of oscillation parameters with star clusters of different 
densities and masses.

We see 3 possible evolution trends:
\begin{itemize}
\item  Small, low-mass clusters stay bound and move with the filament, if the filament is moving slowly.
\item Fast-moving filaments can eject and destroy star clusters
\item Large, high-mass clusters stay bound and the filament just moves through, pumping energy into the cluster.
\end{itemize}

The fate of the cluster in an oscillating potential is decided quickly.
By the time the filament reaches its maximum amplitude, any star
that manages to stay inside the cluster or the gas filament will likely stay
there for the rest of the simulation time. 
A cluster like the ONC might be ejected or stay inside the filament, but these simulations suggest that it will endure the stirring by the filament and it will not be destroyed.

The physics behind the origin of the sharp transition, 
shown in Figure \ref{fig:fractions} (top right panel) is
still under 
investigation, but we believe that it might be  related to 
the mass ratio  between the cluster and the gas filament.


\begin{acknowledgement}
DM gratefully acknowledges support from the Chilean BASAL Centro de Excelencia en Astrofisica y Tecnologias Afines (CATA) grant AFB-170002 and Fondecyt regular No. 1180291. 
MF acknowledges support through Fondecyt regular 1180291, Basal AFB-170002 and PII20150171. 
AS acknowledges funding through Fondecyt Regular (project code
1180350), ''Concurso Proyectos Internacionales de Investigaci\'on''
(project code PII20150171), and Chilean Centro de Excelencia en
Astrof\'isica y Tecnolog\'ias Afines (CATA) BASAL grant AFB-170002.
\end{acknowledgement}


\bibliographystyle{baaa}
\small
\bibliography{biblio}
 
\end{document}